\documentclass[preprint]{aastex}
\usepackage{emulateapj5}
\usepackage{graphicx}

\begin{document}

\title{Are there field-free gaps near {\Large $\tau$}$=1$ in sunspot penumbrae ?}

\author{J.M.~Borrero}
\affil{High Altitude Observatory (NCAR), 3080 Center Green Dr. CG-1, Boulder, CO 80301, USA}
\email{borrero@ucar.edu}
\and
\vspace{-0.5cm}
\author{S.K.~Solanki}
\affil{Max-Planck-Institut f\"ur Sonnensystemforschung, Max-Planck-Strasse 2, 37191 Katlenburg-Lindau, Germany}
\email{solanki@mps.mpg.de}

\begin{abstract}
{The vertical stratification of the magnetic field strength in sunspot penumbrae is investigated by means
of spectropolarimetric observations at high spatial resolution from the Hinode spacecraft. Assuming that
the magnetic field changes linearly with optical depth we find that, in
those regions where the magnetic field is more inclined and the Evershed flow is strongest (penumbral intraspines),
the magnetic field can either increase or decrease with depth. Allowing more degrees of freedom to
the magnetic field stratification reveals that the magnetic field initially decreases 
from $\log\tau_5 = -3$ until $\log\tau_5 \simeq -1.0$, but increases again below that. The
presence of strong magnetic fields near the continuum is at odds with the existence of regions void of magnetic 
fields at, or right below, the $\tau_5=1$ level in the penumbra. However, they are compatible with the presence of 
a horizontal flux-tube-like field embedded in a magnetic atmosphere.}
\end{abstract}


\shorttitle{Field-free gaps near {\Large $\tau$}$=1$ in sunspot penumbrae ?}
\shortauthors{BORRERO \& SOLANKI}
\maketitle

\def\nn{{\bf \nabla}}
\def\cro{\times}
\def\er{{\bf{\rm e_{\rm r}}}}
\def\et{{\bf{\rm e_{\rm \theta}}}}
\def\ex{{\bf{\rm e_{\rm x}}}}
\def\ey{{\bf{\rm e_{\rm y}}}}
\def\ez{{\bf{\rm e_{\rm z}}}}
\def\l{{\bf{\rm l}}}
\def\nx{\mathcal{N}}
\def\sx{\mathcal{S}}
\def\rx{\mathcal{R}}

\section{Introduction}

It is now widely accepted that the horizontal structure of the sunspot penumbra is composed of two
magnetic components (Solanki 2003; Bellot Rubio 2003). One of them possesses a somewhat inclined ($\simeq 40-50^{\circ}$
with respect to the vertical direction to the solar surface) and strong ($\sim 2000$ G) magnetic field,
whereas the other is characterized by a weaker and more horizontal one (Lites et al. 1993; R\"uedi et al. 1998;
Bellot Rubio et al. 2004; Borrero et al. 2004, 2005). Traditionally, these two magnetic 
components have been identified with a horizontal flux tube, that carries the Evershed flow, and is embedded in a more vertical 
background magnetic field: {\it uncombed} model (Solanki \& Montavon 1993; Schlichenmaier et al. 1998; Borrero 2007).
Recently, this view has been challenged by Spruit \& Scharmer (2006) and Scharmer \& Spruit (2006), 
who propose instead that the penumbra is formed by magnetic field-free plumes (connected to the underlying convection
zone) that pierce the penumbral magnetic field from beneath. This is the so-called  {\it gappy} penumbral model.

So long as these two different magnetic structures (weak/horizontal and strong/vertical) have remained
spatially (horizontally) unresolved, distinguishing between the uncombed and gappy penumbral scenarios
has not been possible. However, with the new spectropolarimeter on board of the Japanese spacecraft 
Hinode (Kosugi et al. 2007; Shimizu et al. 2007) it is now possible to obtain high spatial resolution 
($\simeq 0.32$") observations of the sunspot penumbra. This could be sufficient to distinguish
between the uncombed and gappy models, since they both postulate the existence of flux tubes or
field-free gaps that are about 200-300 km in diameter (Mart{\'\i}nez Pillet 2000; Spruit \& Scharmer
2006). This feature is particularly interesting, because the {\it uncombed} and {\it gappy} models predict
very different vertical stratifications in the magnetic field strength across the weak/horizontal magnetic 
field component: which is identified with an embedded flux tube in the uncombed model, but with a 
field-free gap in the gappy model. In the latter, the magnetic field decreases monotonically with depth,
whereas the former possesses a magnetic field that decreases with depth only initially, since
once the boundary of the flux tube is reached, the magnetic field can either decrease or increase
depending upon the strength of the magnetic field inside the tube.

In this paper we will focus on obtaining the vertical stratification of the magnetic field for penumbral filaments
(where the magnetic field is more horizontal and weaker) using high spatial resolution spectropolarimetric observations
from Hinode, in order to establish which penumbral model is more realistic. The observations are described in Section 2. 
Section 3 describes our data analysis and results from our inversion technique.  Section 4 compares our findings
with the predictions made by the uncombed and gappy penumbral models. In Section 5 we make a thourough investigation
of the effects of the scattered light. Finally, Section 6 summarizes our findings.

\section{Observations}

On May 3rd 2007, between 10:15 and 11:40 am UT, the active region AR 10953 was mapped using the spectropolarimeter of
the Solar Optical Telescope on-board of the Hinode spacecraft (Lites et al. 2001). The active region 
was located at a heliocentric angle of $\theta=19.2^{\circ}$.
It was scanned in a thousand steps, with a step width of 0.148" and a
slit width of 0.158". The spectropolarimeter recorded the full Stokes vector ($I$, $Q$, $U$ and $V$)
of the pair of neutral iron lines at 630 nm with a spectral sampling of 21.53 m\AA. 
The integration time was 4.8 seconds, resulting in an approximate noise level of $1.2 \times 10^{-3}$ (in units 
of the normalized continuum intensity). In the absence of the telluric oxygen lines we proceeded with two different wavelength calibration
methods that were cross-checked for consistency. The first method was obtained by matching the average
quiet Sun profile with the FTS spectrum, whereas the second calibration assumes that the average umbral
profile exhibits no velocities.

A map of the continuum intensity at 630 nm of the scanned region is shown in Figure 1. The white 
arrow indicates the direction of the center of the solar disk. The penumbra on the center 
side is heavily distorted and therefore left out from our analysis.
On the limb side the penumbra is more uniform, with radially aligned filaments. The
region enclosed by the white rectangle has been chosen for our study. This sunspot has negative polarity 
(magnetic field in the umbra points towards the solar interior), however the results presented hereafter are shown, 
in order to facilitate the interpretation, as if the sunspot had positive polarity.

\begin{center}
\includegraphics[width=8cm]{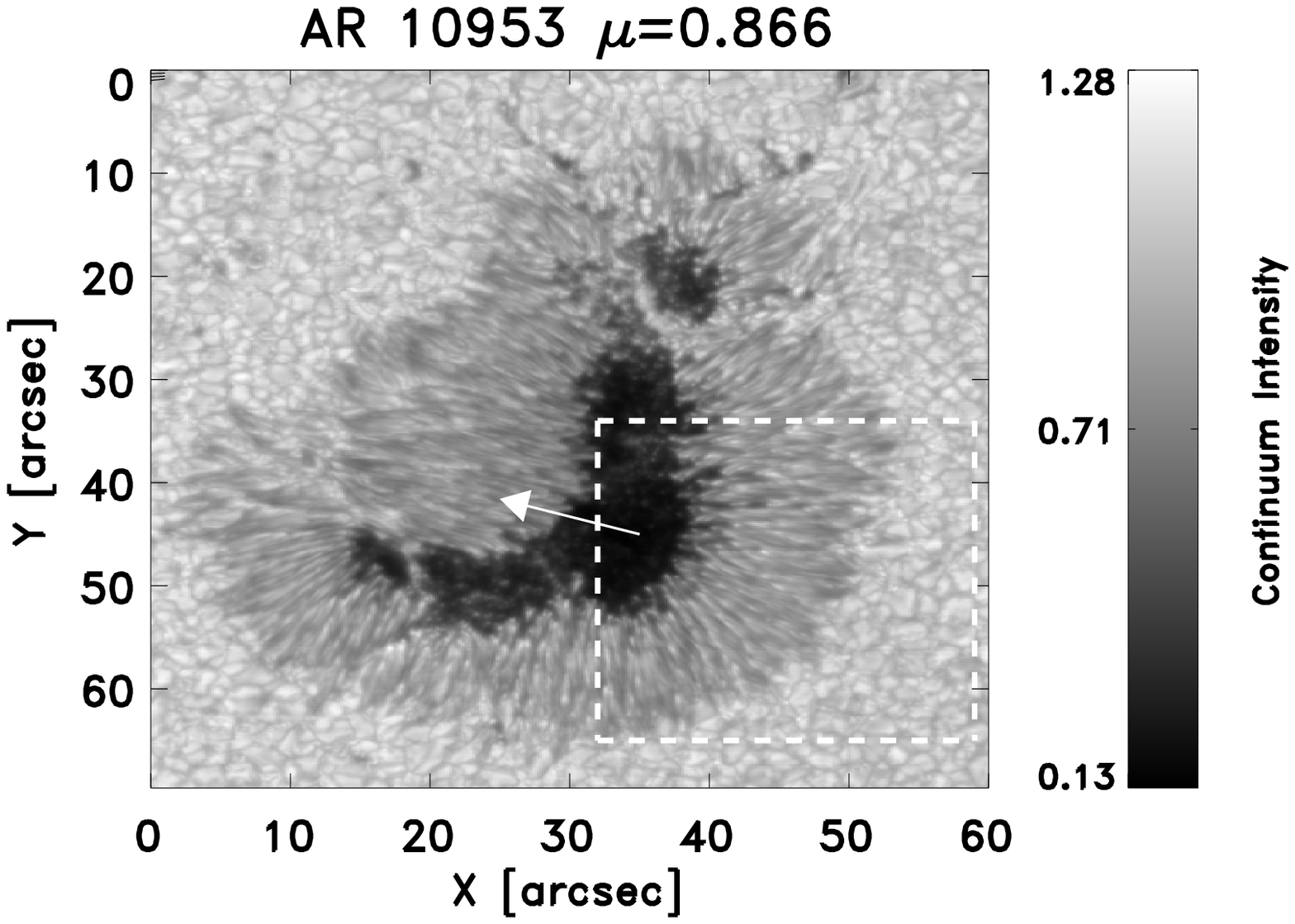}
\figcaption{Continuum intensity map at 630 nm of AR 10953. This sunspot
was observed using Hinode's spectropolarimeter on the 3rd of May, 2007 at an 
heliocentric angle of $\theta = 19^{\circ}$. The white arrow points towards the center
of the solar disk. The white rectangle limits the region chosen for our study. It lies
on the limb-side around the line-of-symmetry.}
\end{center}

\section{Data analysis and Results}%

We have applied the SIR inversion code (Ruiz Cobo \& Del Toro Iniesta 1992) to
our spectropolarimetric observations to retrieve the physical properties of the solar atmosphere.
This code allows all relevant physical parameters to be a generic function
of the optical depth: $B(\tau)$, $\gamma(\tau)$, $\phi(\tau)$, $V_{\rm los}(\tau)$, etc. 
In addition, a depth-dependent temperature stratification $T(\tau)$ models the 
atmospheric thermodynamics under local thermodynamic equilibrium (LTE) conditions. SIR retrieves the values of the parameters at a number of optical depth 
points called {\it nodes}. The final stratification is obtained by interpolating splines across those nodes.
Note however, that SIR employs equivalent response functions (Del Toro Iniesta 2003), which ensures sensitivity
to the atmospheric layers located between nodes. Each node represents a free parameter in the inversion. 
In our investigation we will employ increasingly complex models (i.e.: more free parameters) according to the amount of information we hope
to extract from the profiles.

Given the high spatial resolution of Hinode's observations, we will consider only one magnetic 
component. A non-magnetic component is also considered to account for the scattered light. 
In this section, the scattered light profile is obtained by averaging the intensity profiles of those pixels 
with polarization signals below the noise level (quiet Sun granulation around the sunspot). 
The same scattered light profile is used in the inversion of all pixels. In Sect.~5 we make a thourough
analysis of the effects that different treatments for the scattered light have on our results.
Note that using one single magnetic component is equivalent to assuming that the penumbral
structure is horizontally resolved. This is clearly not the case if we look into continuum images
at even higher spatial resolution (Scharmer et al. 2002). However, our assumption would still be valid if the 
(unresolved) variations of the magnetic field inside the weak/horizontal magnetic component are much smaller 
than the differences between the weak/horizontal and strong/vertical components. Since the former remain unresolved, 
we cannot assess the validity of this assumption. This question should be addressed as better spectropolarimetric 
observations become available.

\subsection{1-node inversion and intraspine selection}

In order to locate the intraspinal pixels we have carried out a first inversion where all physical parameters, 
with the exception of the temperature, are constant with optical depth. We therefore have one single node for  $B(\tau)$, $\gamma(\tau)$, $\phi(\tau)$, 
$V_{\rm los}(\tau)$. To account for unresolved velocity fields, we also consider depth-independent
micro and macroturbulent velocities: $V_{\rm mic}$ and $V_{\rm mac}$. Another
free parameter, $\alpha_{\rm qs}$, represents the fraction of the observed intensity, Stokes $I$, that
corresponds to scattered light. Finally, three nodes are given to the temperature $T(\tau)$.
In total, this first inversion has 10 free parameters. Since $B$, $\gamma$, $\phi$ and $V_{\rm los}$ are constant
with optical depth, the retrieved values indicate some kind of average over the region where the spectral lines
are formed. Westendorp Plaza et al. (1998,2001) studied this issue in detail and found that the largest contribution
for this pair of Fe I lines (Sect.~2) comes from $\log\tau_5 \simeq -1.5$.

Figure 2 displays the resulting values for the line-of-sight velocity and magnetic field vector in the selected
box in Fig.~1. Regions of weak, $B < 1300$ G, and highly inclined, $\gamma > 80^{\circ}$, 
magnetic field can be clearly distinguished in this figure. They are also characterized by
the presence of large red-shifted velocities (Evershed flow). These are the so-called penumbral intraspines, and therefore the
most likely locations where field-free gaps or horizontal flux-tubes can be found. Also visible are structures characterized by a stronger 
and more vertical magnetic field, as well as by a strongly reduced Evershed flow. These are usually referred to as spines.
Spines and intraspines are also seen at moderate ($\sim$ 1") spatial resolution (Lites et al. 1993; Stanchfield et al. 1997; 
Mathew et al. 2003) but the associated changes in their properties (field strength, inclination, etc) are 
larger if observed at high spatial resolution (Bello Gonz\'alez et al. 2005; Langhans et al. 2005).
Bellot Rubio et al. (2004) interprets this result as a consequence of these structures not being spatially
resolved at 1" resolution. As demonstrated by Borrero et al. (2008) they are indeed horizontally resolved
in Hinode observations (0.32").

\begin{figure*}
\begin{center}
\includegraphics[width=14cm]{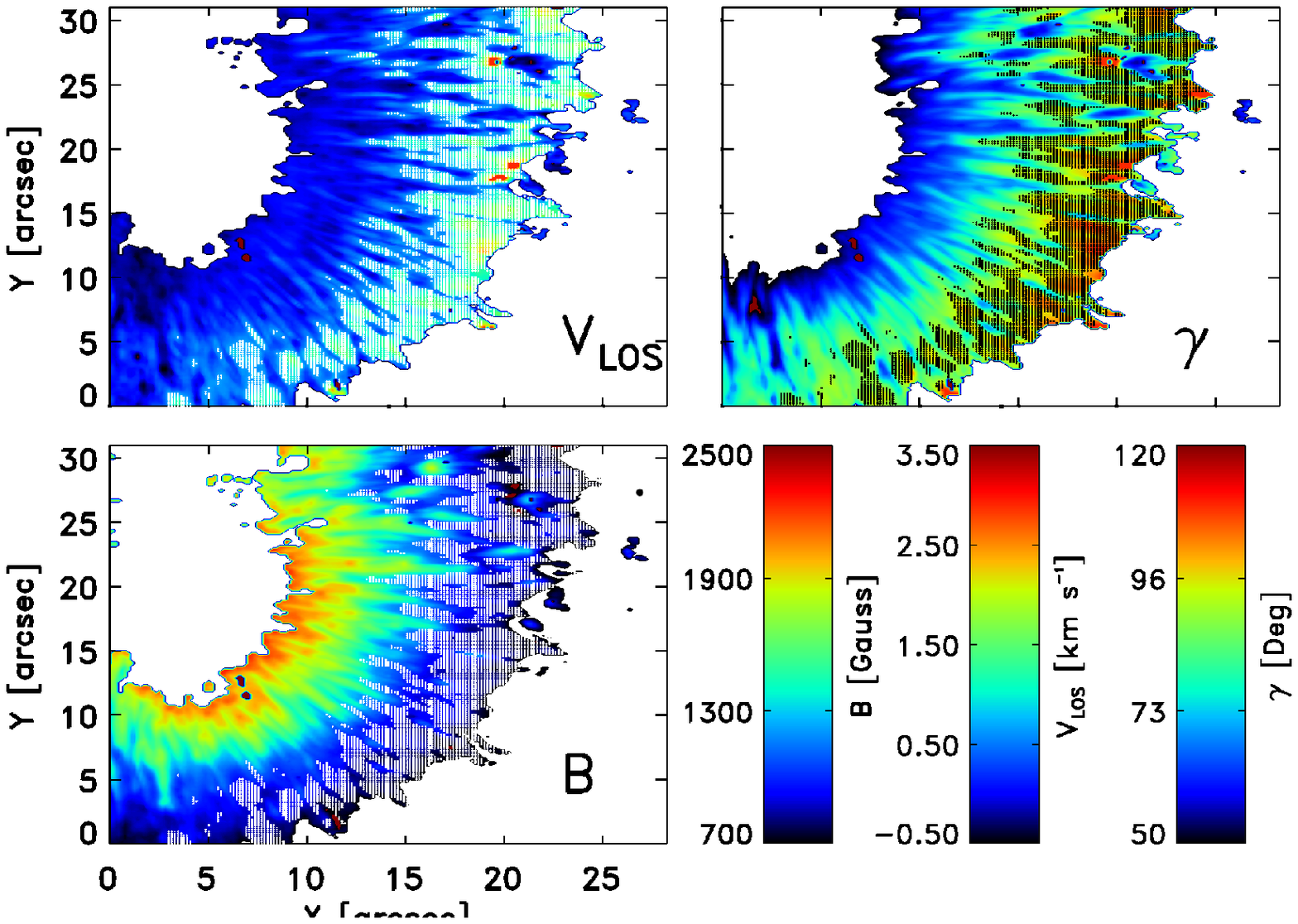} \\
\figcaption{Results from the inversion of the region limited by the square box in Fig.~1. The inversion was
performed assuming that $B$, $\gamma$, $\phi$ and $V_{\rm los}$ are constant with optical depth. The
magnetic field strength is displayed in the lower-left panel, inclination (upper-right),and
line-of-sight velocity (upper-left). The white and black dots correspond to those pixels where
the location of horizontal flux tubes or field-free gaps are suspected (see text for details). There are
7520 of them: 39 \% of all penumbral pixels in this figure.}
\end{center}
\end{figure*}

\begin{figure*}
\begin{center}
\includegraphics[width=14cm]{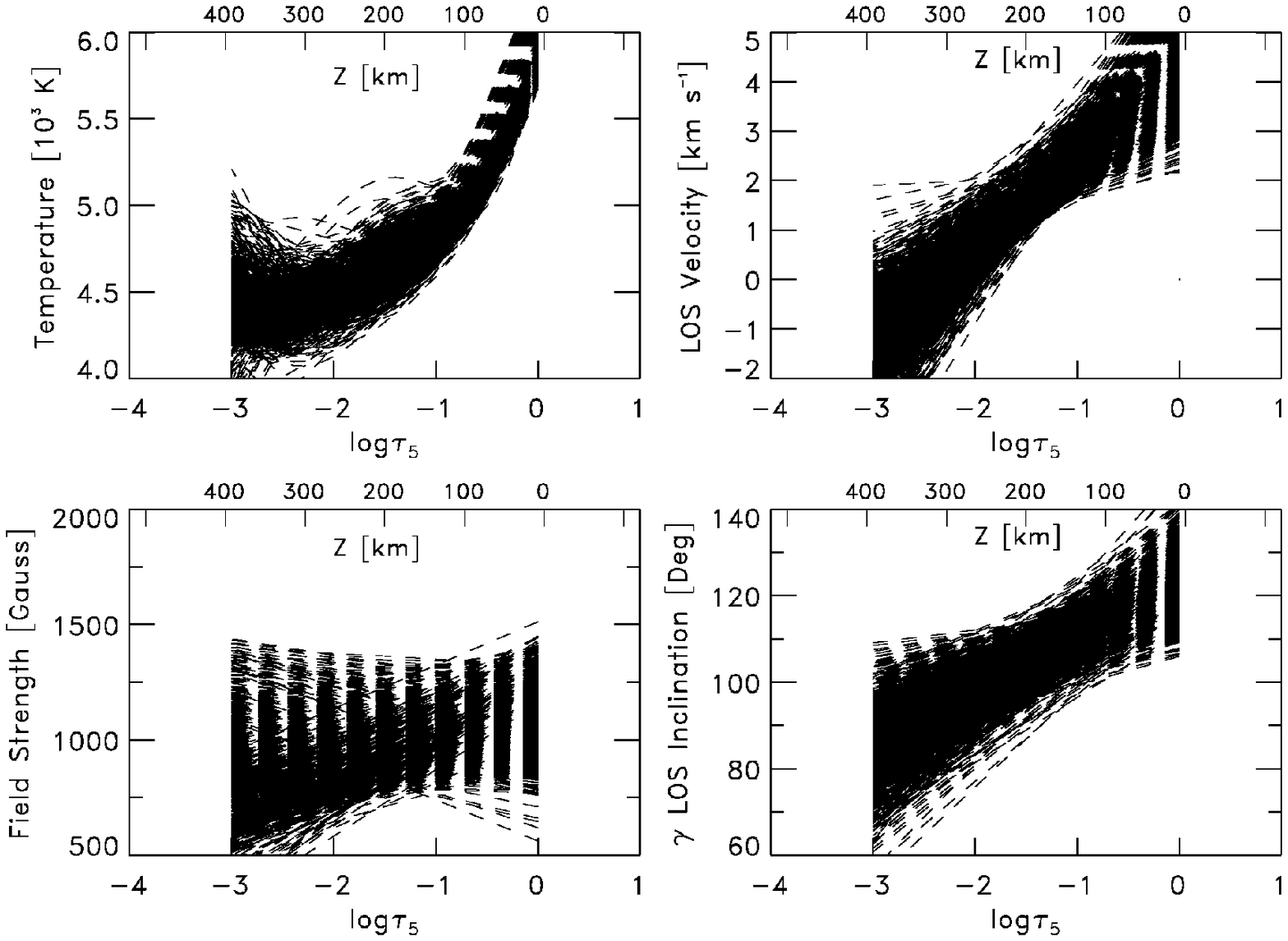} \\
\figcaption{Temperature (upper-left), line-of-sight velocity (upper-right), magnetic field strength (lower-left),
and magnetic field zenith angle (lower-right) as a function of the optical depth $\log\tau_5$, obtained from the
2-node inversion of the intraspinal pixels selected in Fig.~2. Approximate height scale, Z, computed assuming vertical hydrostatic 
equilibrium, is also indicated.}
\end{center}
\end{figure*}

In Figure 2 we also indicate with black and white dots a large number (total of 7520) of intraspinal pixels.
They have been found as those satisfying the following conditions: $B > 700$ G, and $V_{\rm los} \in [1.0,3.0]$ km s$^{-1}$.
Since the main difference between spines and intraspines is the presence of a strong Evershed flow, we use $V_{\rm los}$ to distinguish among them.
However, we do not consider the few pixels where $V_{\rm los} > 3$ km s$^{-1}$, since they usually present extremely 
abnormal Stokes V profiles, usually a sign of the existence of horizontally unresolved structure. We do not constrain the values of the magnetic 
field inclination and strength (here we use only a lower limit to avoid taking pixels outside the visible boundary of the sunspot) because the 
magnetic properties of the spines in the outer penumbra are very similar to those of the intraspines in the inner penumbra. 
The final selected pixels represent about 39 \% of all penumbral pixels in Figure 2. Note that they are mostly located in the middle and outer penumbra:
$r/R_s > 0.5$ ($R_s$ being the sunspot radius; umbral-penumbral boundary is located at $r/R_s > 0.25$). Note also that, even though
we have not constrained the values of the magnetic field strength an inclination, all intraspinal pixels are located in
regions where the magnetic field is highly inclined and weak.

\subsection{2-node inversion of individual profiles}

In order to investigate the depth variation of the physical parameters in intraspines, we performed a renewed inversion of the pixels selected
 in Fig.~2, where we now allow for two nodes in $B(\tau)$,  $\gamma(\tau)$, $\phi(\tau)$, $V_{\rm los}(\tau)$ (linear variations with 
optical depth). The total number of free parameters is now 14. Results from this new inversion are presented in Figure 3: $T(\tau)$ (upper-left),
$V_{\rm los}(\tau)$ (upper-right), $B(\tau)$ (lower-left), and $\gamma(\tau)$\ (lower-right). All inverted pixels display similar 
stratifications of $V_{\rm los}(\tau)$  and $\gamma(\tau)$: both increase monotonically with optical depth: $\partial V_{\rm los}/\partial \tau$, 
$\partial \gamma/\partial \tau>0$. The magnetic field strength $B(\tau)$, however, can either increase (in 66 \% of inverted pixels: 4971) 
or decrease (34 \%; 2249 pixels) with optical depth. In either case, the retrieved gradient is relatively small: $|dB/dz| \le 1.5$ Gauss km$^{-1}$.
An important feature to note is that pixels displaying a decreasing magnetic field towards deeper layers, $dB/ d \tau < 0$,
are mostly located in the inner penumbra, whereas pixels showing $dB/ d \tau > 0$ are mostly found in the outer penumbra (see Figure 4).

\begin{center}
\includegraphics[width=8cm]{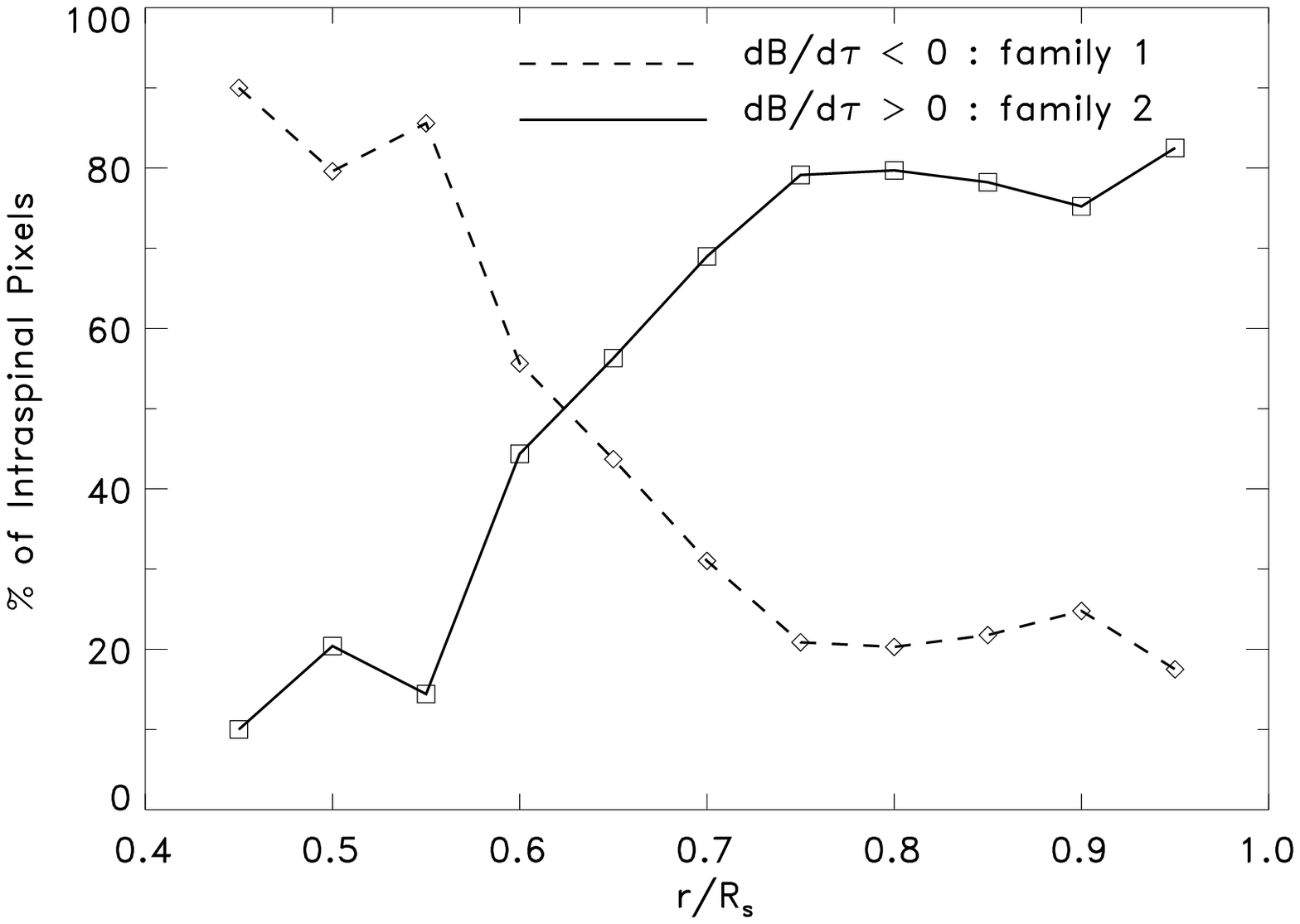}
\figcaption{Percentage of the total selected pixels that, at each radial distance from sunspot center, show 
a magnetic field that increases ($dB/ d \tau > 0$; solid line) or decreases ($dB/ d \tau > 0$; dashed line)
towards the solar interior.}
\end{center}

\subsection{4-node inversion of individual profiles}

We now perform a more  complex inversion of the same pixels as in Sect.~3.2. In this case
we allow for 4 nodes in $B(\tau)$, $\gamma(\tau)$, $\phi(\tau)$, and $V_{\rm los}(\tau)$.
These nodes are located at optical depth positions: $\log\tau_5=[-3.2,-1.8,-0.4,1]$.
In total, this new inversion has 22 free parameters. Figure 5 shows the results from the 4-node 
inversion of the 7250 intraspinal pixels selected in Sect.~3.1. The stratifications are
very similar to those already obtained through the 2-node inversion (see Fig.~3).
The larger scatter (pixel-to-pixel variations) in the 4-node inversion is due to the larger amount of 
free parameters, which are more weakly constrained by the observations. 

Since now we allow for 4 nodes to the stratification of the magnetic field strength it is not 
easy to classify our results between those where the magnetic field increases or 
decreases with optical depth. To showcase the differences between the possible stratifications
we have taken separately those pixels where, in the 2-node inversion, showed  $dB/ d \tau < 0$
(family 1) or $dB/ d \tau > 0$ (family 2) and obtained the averaged stratification for the 2 and
4-node inversion. Results for family 1 and 2 are presented in Figures 6 and 7 respectively.

\begin{figure*}
\begin{center}
\includegraphics[width=14cm]{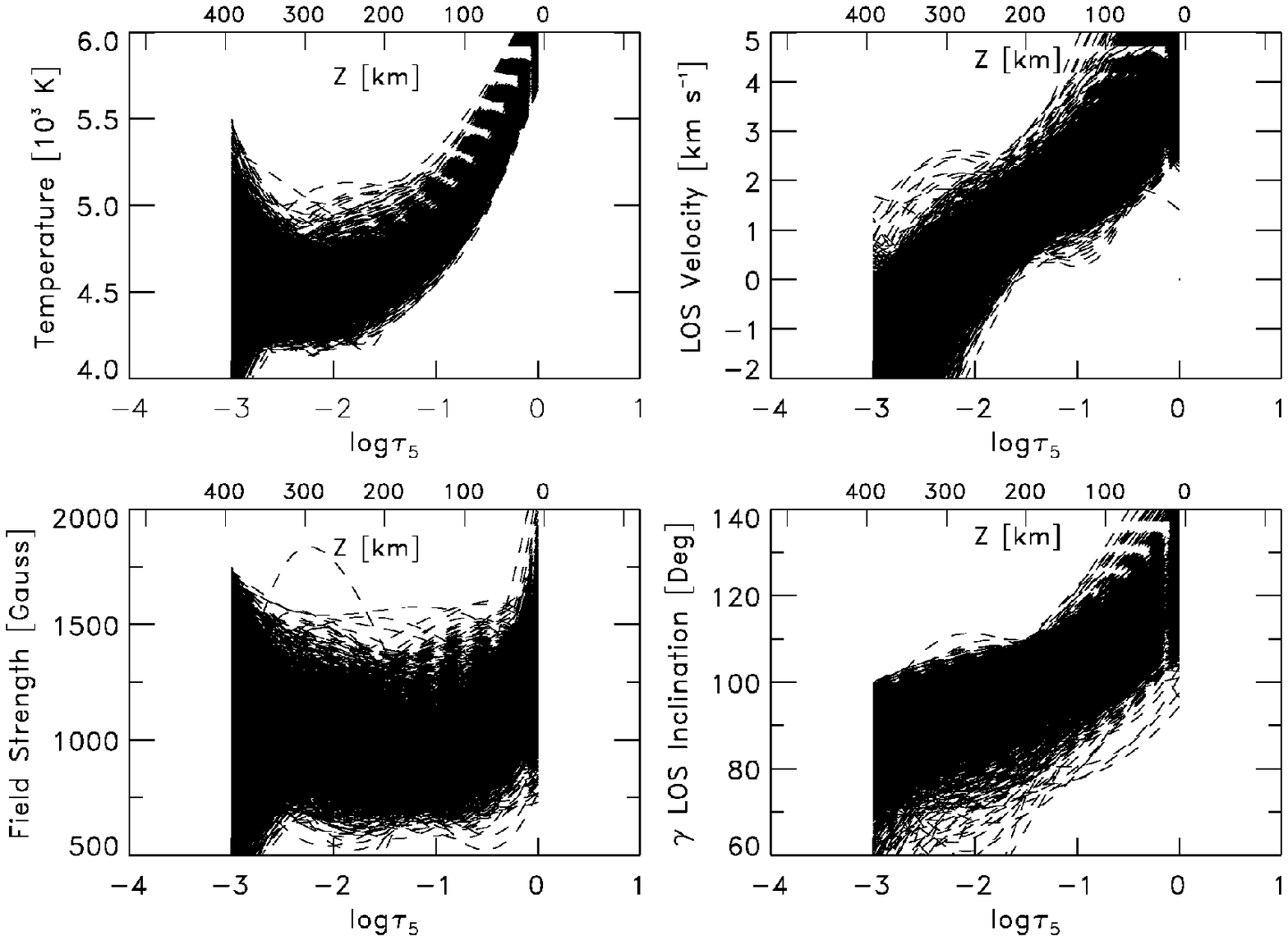} \\
\figcaption{Same as Figure 3 but for the 4-node inversion.}
\end{center}
\end{figure*}

Interestingly, the magnetic field strength, that in the 2-node inversion showed different gradients, now
shows an initial decrease up  to $\log\tau_5 \simeq -1$ (approximately 100 km above the continuum level), where it starts to
 increase again towards deep layers. This happens for both families of magnetic structures, and thus could indicate that they
are indeed closely related. A closer look reveals that both families possess a similar magnetic field strength in deep layers: 
$B(\tau=1) \simeq 1250$ G, but slightly different higher up: $B(\tau=10^{-3}) \simeq 1200$ G (family 1; Fig.~6) and 
$B(\tau=10^{-3}) \simeq 900$ G (family 2; Fig.~7). This effect explains why the 2-node inversion (Sect.~3.2) retrieves different 
overall gradients for the magnetic field strength: it is due to a large variation in the magnetic field at around 
$\tau_5 \simeq 10^{-3}$ since the magnetic field deeper down is basically the same in both cases.

To further confirm these results we have repeated our 4-node inversion with the nodes located at slightly different positions.
SIR always places 2 nodes at the uppermost and deepest $\tau$-locations of the discretized atmosphere, while spreading the rest
equidistantly in between. Therefore, to keep the same number of nodes and, at the same time, change their $\tau$-positions
we must change the initial and last $\tau$-points of the atmosphere. In our first set of inversions the atmosphere is discretized
between $\log\tau_5=[-3.2,1]$. Changing this to $\log\tau_5=[-3.0,1.2]$ and $\log\tau_5=[-3.7,0.5]$ would position the 4 nodes
at $[-3.0,-1.6,-0.2,1.2]$ and $[-3.7,-2.3,-0.9,0.5]$, respectively. We have inverted all pixels again in these two cases and confirmed
that our results (Fig.~5,6 and 7) do not change. This is due to the use that SIR makes of equivalent response functions (see Section~3). 

\begin{figure*}
\begin{center}
\includegraphics[width=14cm]{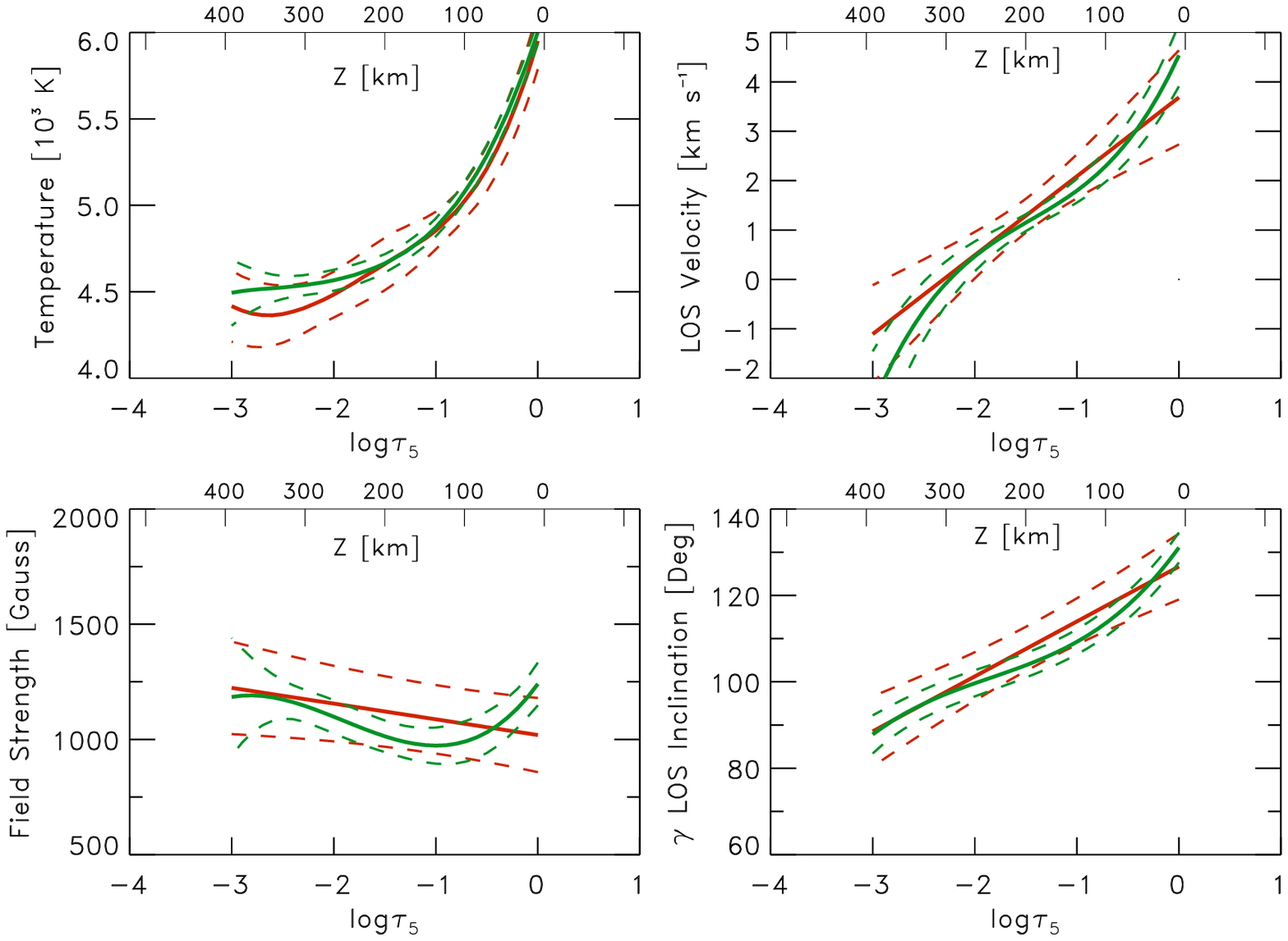}
\figcaption{Temperature (top left), line-of-sight velocity (top right), magnetic field strength (lower left)
, and magnetic field inclination (lower right) as a function of the optical depth. Red indicates the average stratification
obtained from the individual 2-node inversion of the 2549 profiles belonging to family 1: $dB/ d\tau < 0$ (taken from Fig.~3). Green 
shows the average stratification obtained from the individual 4-node inversion of the Stokes vector of the same pixels.}
\end{center}
\end{figure*}

\begin{figure*}
\begin{center}
\includegraphics[width=14cm]{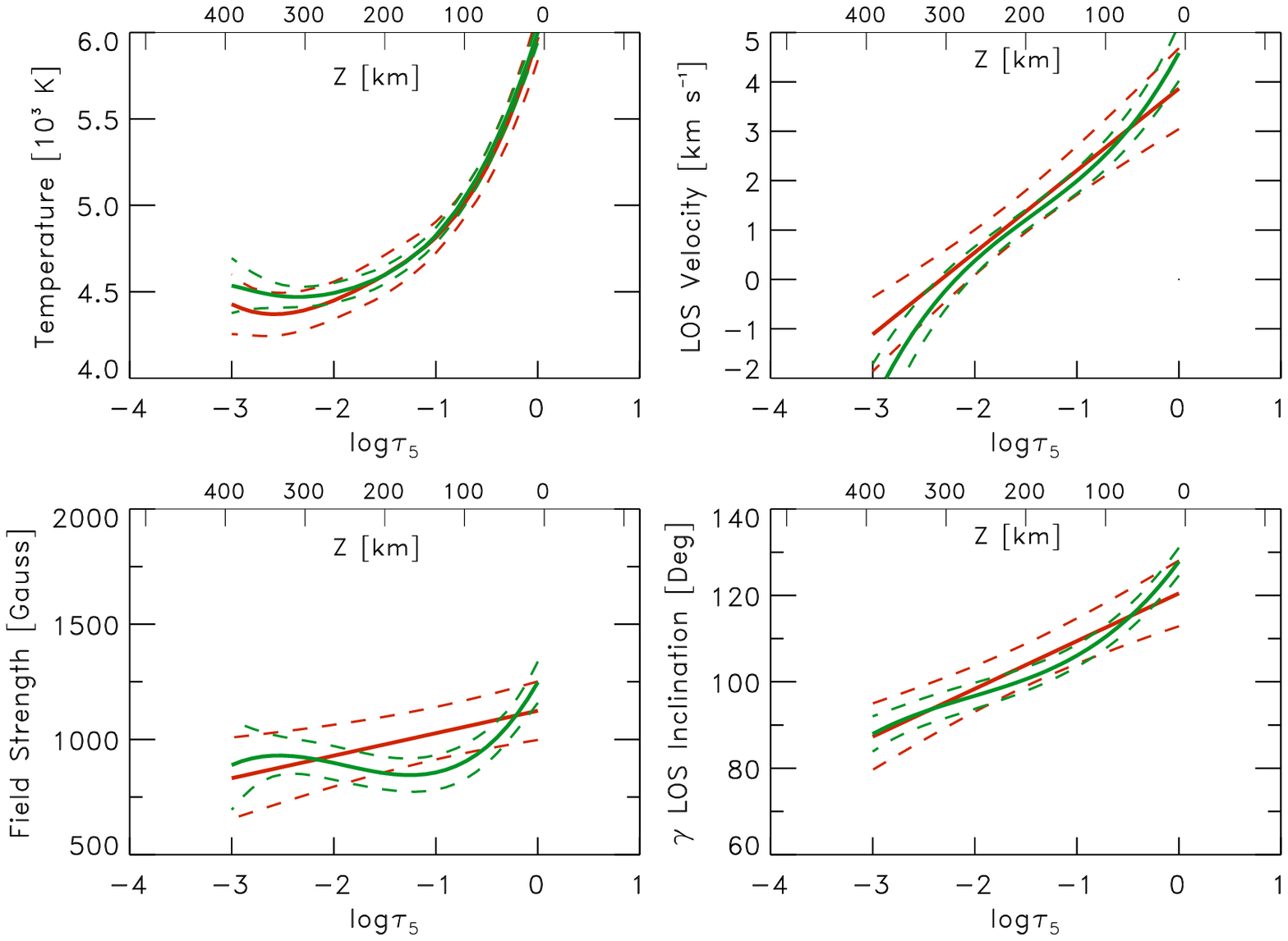}
\figcaption{Same as Figure 6 but for the 4971 pixels belonging to family 2: $dB/ d\tau > 0$.}
\end{center}
\end{figure*}

\section{Discussion}%

If we consider a ray passing through the center of an intraspine, the {\it gappy} and {\it uncombed} penumbral models 
predict a very similar stratification of the magnetic field strength above the field-free gap or flux tube, but very different 
ones inside them. Figure 8 illustrates some possible stratifications predicted by these two models, where the
upper boundary of the field-free gap and flux tube is located at $z=0$. Above $z=0$ they both share the same stratification for
the surrounding magnetic atmosphere.  Here we present two examples, one where the surrounding magnetic field is weak: $B_{\rm surr}=1000$ (dashed
line; meant to represent the outer penumbra, $r/R_s=0.8$), and another case where the surrounding field is stronger (solid
line): $B_{\rm surr}=1500$ (meant to represent to inner penumbra, $r/R_s=0.4$). Note that the magnetic field strength in the surrounding 
atmosphere decreases towards deeper layers. This is due to the fact that the vertical component of the surrounding field must vanish
(or nearly vanish in the case of a cusp-shaped boundary) at the flux tube's or gap's boundary. These two examples are
actual solutions of analytical models (Fig.~5 in Spruit \& Scharmer 2006; Fig.~3 in Scharmer \& Spruit 2006; see also 
Eqs.~33-34 in Borrero 2007). Below the boundary of the flux tube or field-free gap, $z=0$, both models predict a very different situation. In
the case of the {\it gappy} penumbra this region is void of magnetic fields: $B_{\rm gap} \simeq 0$ (hollow circles). In contrast, the {\it uncombed}
model assumes the existence of a flux tube where the magnetic field is strong $B_{\rm tube}=1250$ G (filled circles).

If we compare Fig.~8 with our 2-node inversion (Fig.~3) of intraspinal pixels we find that, on the one hand, the {\it gappy} penumbral model 
can only explain the slowly decreasing magnetic field, observed for 34 \% of intraspinal pixels (family 1), if the $\tau_5=1$ level is 
formed above the gap's boundary, otherwise a much more sudden drop would be observed (hollow circles in Fig.~8).
On the other hand, this model does not offer any explanation for the 66 \% of the intraspinal pixels that present an
increasing magnetic field strength towards deeper layers (family 2). However, the {\it uncombed} penumbral model can explain both observed 
situations. It all depends on the strength of the flux tube's magnetic field 
as compared to the magnetic field high above it: $B_{\rm surr}$ versus $B_{\rm tube}$. A magnetic field that 
decreases smoothly towards the interior of the photosphere can be explained by a flux tube (of any field strength)
whose upper boundary layer lies below $\tau_5=1$. If the upper boundary is above $\tau_5=1$, it can also be explained with a
magnetic field inside the flux tube that is weaker than the magnetic field a
few hundred kilometers above (solid lines plus filled circles in Fig.~8).  In addition, a magnetic field that increases towards
the interior of the photosphere is compatible with a flux tube with an upper boundary layer
above $\tau_5=1$, and with a stronger magnetic field than the one above (dashed line and filled circles in Fig.~8). 

\begin{center}
\includegraphics[width=8cm]{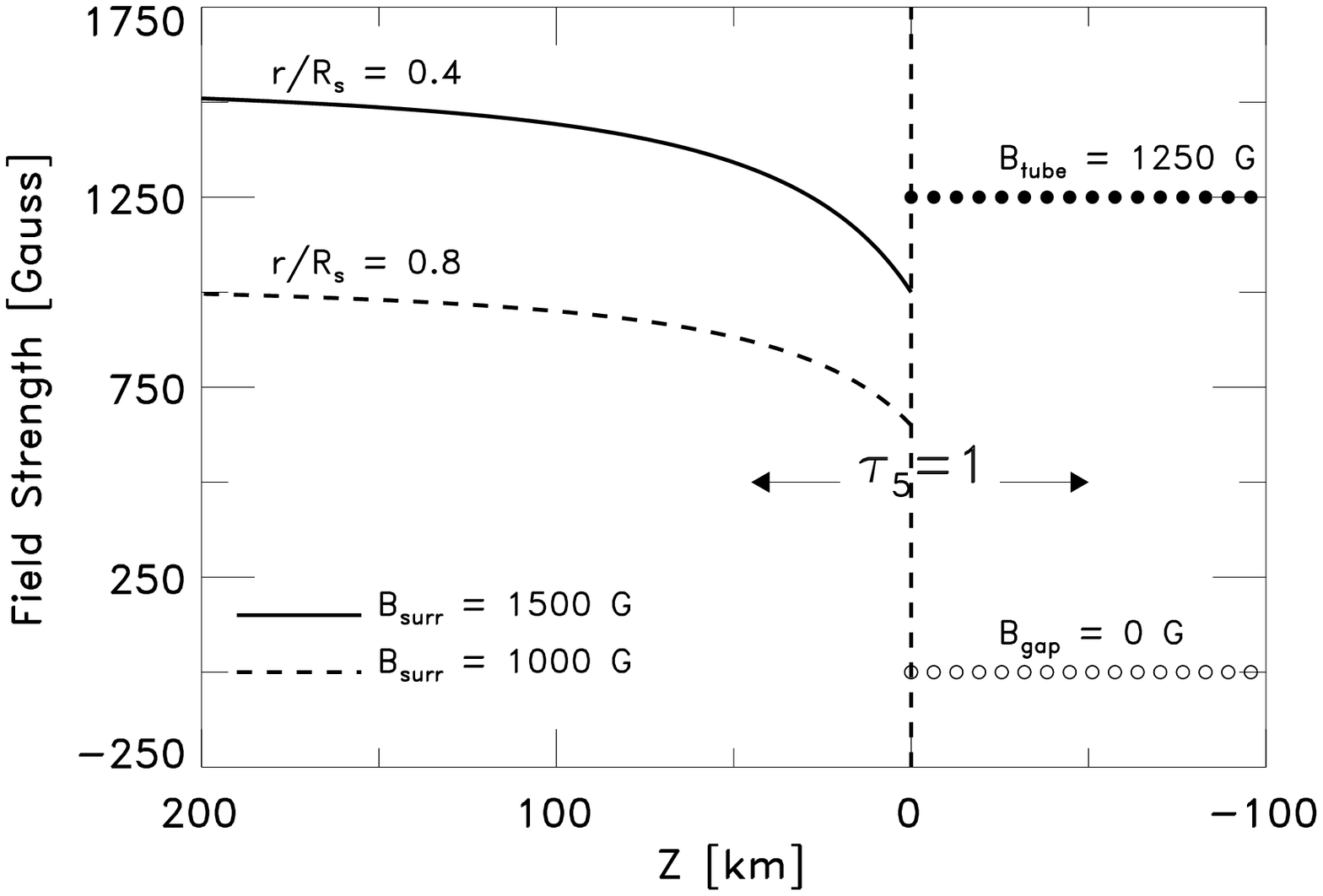}
\figcaption{Vertical variation of the magnetic field strength across the center of a field-free gap (hollow circles) according to 
the gappy penumbral model. Same for a flux tube with a magnetic field strength of 1250 G (filled circles). Note that both models
share the same stratification above the tube's or gap's boundary ($z > 0$). The solid line represents a situation where the
external field is rather strong (inner penumbra), while the dashed line corresponds to the outer penumbra (weak external field).
Also, note that the $\tau_5=1$ level can be shifted horizontally such that the continuum level can be formed above the 
gap/flux-tube or inside them.}
\end{center}

A more complex (4-nodes) inversion of intraspinal profiles indicates that, what appeared as two different families of structures 
using a 2-node inversion, are likely to correspond to one single kind of magnetic structure, where the magnetic field exhibits 
an initial decrease between $\log\tau_5 \in [-3,-1.0]$, but increases between $\log\tau_5 \in [-1.0,0]$ (see Figs.~6-7). 
While the {\it gappy} model offers no explanation for this effect,
it can indeed be explained by the {\it uncombed} penumbral model, by means of a magnetic field whose strength decreases initially but
increases once the line-of-sight crosses the flux tube's boundary (see Fig.~8). Furthermore, although intraspinal families 1 and 2 appear 
to be the equivalent in the 4-node inversion, they still present a subtle yet important difference: family 1 
(more commonly found in the inner penumbra; see Fig.~4) displays a much stronger initial decrease as compared to family 2,
which is usually found in the outer penumbra (compare lower-left panels in Figs.~6 and 7). 

This can be explained, in terms
of the {\it uncombed} model if, at small-intermediate radial distances, the horizontal flux tube possesses a weaker magnetic field than the field in the 
atmosphere in which it is embedded: $B_{\rm tube} < B_{\rm surr}$ at $r/R_s$ small (compare solid line plus filled circles in Fig.~8 with 
green solid in Fig.~6). As we move towards larger radial distances, and assuming that the magnetic field 
inside the flux tube remains constant, the surrounding magnetic field weakens and falls below the flux tube's field strength: 
$B_{\rm tube} > B_{\rm surr}$  at $r/R_s$ large (compare dashed line plus filled circles in Fig.~8 with solid green in Fig.~7). Note that 
the assumption that the magnetic field in the flux tube remains constant is in agreement with a surrounding magnetic field whose strength decays 
much more rapidly towards the outer penumbra than inside the flux tube (see Fig.~4 in Borrero et al. 2004; Fig.~6 in Borrero et al. 2005 and Fig.~4 in 
Borrero et al. 2006). Here we find that this known feature of the penumbral intraspines helps to explain, within the frame of the uncombed model, 
differences in the stratification in the magnetic field strength across instrapines at different radial distances, as deduced from high resolution 
spectropolarimetric observations.

\section{Scattered light considerations}%

One of the most critical issues in the inversion of spectropolarimetric data is the treatment of the scattered 
light. In order to properly model its contribution, detailed measurements of the telescope's PSF are needed.  
Since these are not usually available, the scattered light is often 
 treated as a non-polarized contribution to the total observed light (see Sect.~3). In our study this is particularly important 
because one of the models under study ({\it gappy} model) postulates the existence of field-free regions around the $\tau_5=1$ level 
in the penumbra. These regions will naturally produce a non-polarized contribution to the total observed Stokes vector. 
Therefore, there is a potential risk of not detecting the field-free gaps due to an incorrect treatment of the scattered light. 

\begin{center}
\includegraphics[width=9cm]{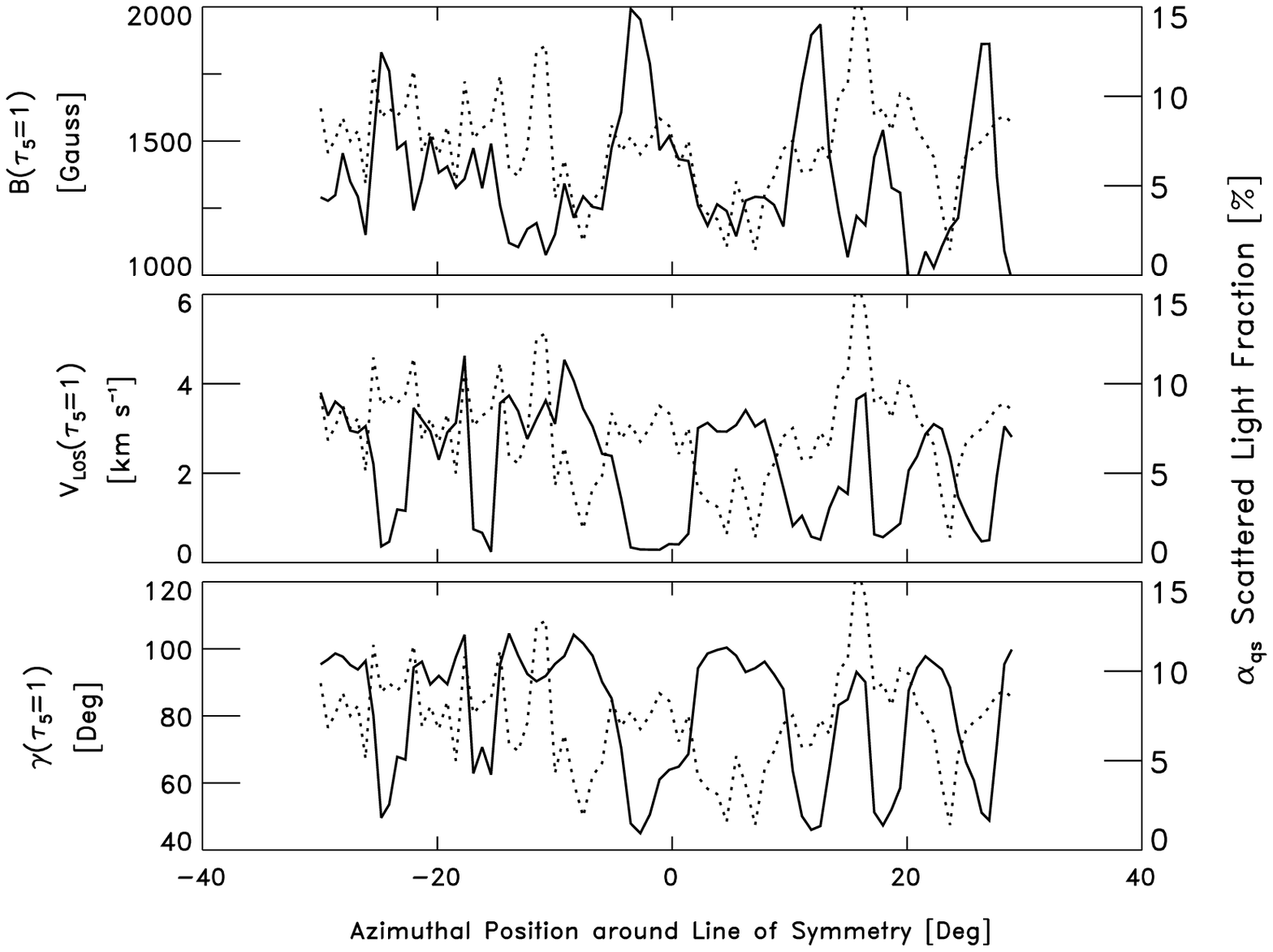} \\
\figcaption{Azimuthal variation of the magnetic field strength (top), line-of-sight velocity (middle), and
inclination of the magnetic field (bottom) at $\tau_5=1$, for a cut at a radial distance $r/R_s=0.75$ in Fig.~2.
The origin of the azimuth angle (abscissa) corresponds to the line of symmetry of the sunspot, indicated by the arrow in Fig.~1. The
scattered light filling factor, $\alpha_{\rm qs}$, is also plotted in all three panels (dashed lines).}
\end{center}

If our inversions are affected by this degeneracy between scattered light and field-free gaps, it is expected
that those pixels where the intraspines are located show larger values for the amount of scattered light retrieved by
the inversion ($\alpha_{\rm qs}$). To study this possibility we have plotted in Figure~9  the variations of
the magnetic field strength (top panel), line-of-sight velocity (middle panel) and inclination angle (bottom)
along an azimuthal cut at $r/R_s=0.75$. Other azimuthal cuts at different radial distances
show very similar behaviors. The values are taken at an optical depth of $\tau_5=1$
from the 2-node inversion in Sect.~3.2. This plot includes, not only those pixels selected in Fig.~2 as intraspines,
but all of them. Therefore regions where the magnitude of Evershed flow is reduced and the magnetic field is more
vertical and strong (penumbral spines) are also visible. All three panels also show the amount of
scattered light $\alpha_{\rm qs}$ (dashed lines). There is no particular correlation between 
the location of penumbral intraspines (high velocities, weak and very inclined fields) and the regions where $\alpha_{\rm qs}$
is largest. Similar variations are observed if we plot the values of the magnetic field strength and inclination, and line-of-sight velocity,
at an optical depth of $\tau_5=10^{-2}$. This rules out the possibility that our inversions do not show field-free regions, in the deep photospheric
layers, where intraspines are located at the expense of an enhanced scattered light contribution.

Recently, Orozco Su\'arez et al. (2007a; 2007b) have presented inversions of Stokes spectra measured
using Hinode's spectropolarimeter in the quiet Sun. These authors claim that for this instrument it is more appropriate to consider
a local (unpolarized )scattered light profile. This is obtained by averaging the observed Stokes $I$ profiles over a small
region (about 1 arcsec) around the pixel that is being studied. In this case, a different scattered light profile
is used in the inversion of each pixel. This approach can be justified by the fact that the focus of the
Narrow-band filter (NBI) on Hinode is favored when simultaneous observations are carried out with both instruments.
In our inversions, we have however used a global scattered light profile, where we average the Stokes $I$ signal emerging from the quiet Sun region far away from 
the sunspot. In order to test whether our results depend on the use of a different scattered light profile,
we have repeated our 2-node inversion using the same approach as Orozco Su\'arez et al. The results are presented
in Fig.~10 (cf. Fig.~3). The percentage of intraspinal pixels with $dB/d\tau >0$ is even larger than before (90 \%).

Alternatively, Orozco Su\'arez et al.  (2007a; 2007b) point out that the most realistic way to account for the scattered light would be to consider a
local {\bf and} polarized scattered light profile, where not only Stokes $I$ is averaged, but also Stokes $Q$, $U$
and $V$. We have also tested this possibility. Unfortunately, this yields unrealistically high values
for $\alpha_{\rm qs}$ during the inversion: $\alpha_{\rm qs} > 0.9$. This indicates that the inversion code tries to dominantly reproduce the
observed Stokes profiles using the scattered light contribution. This is not surprising since the neighboring Stokes profiles often look 
very similar to those in the pixel under study. Therefore we conclude that this is not
a reasonable approach when inverting sunspot data. We cannot rule out however, that this treatment will work
in quiet Sun regions.

\begin{figure*}
\begin{center}
\includegraphics[width=14cm]{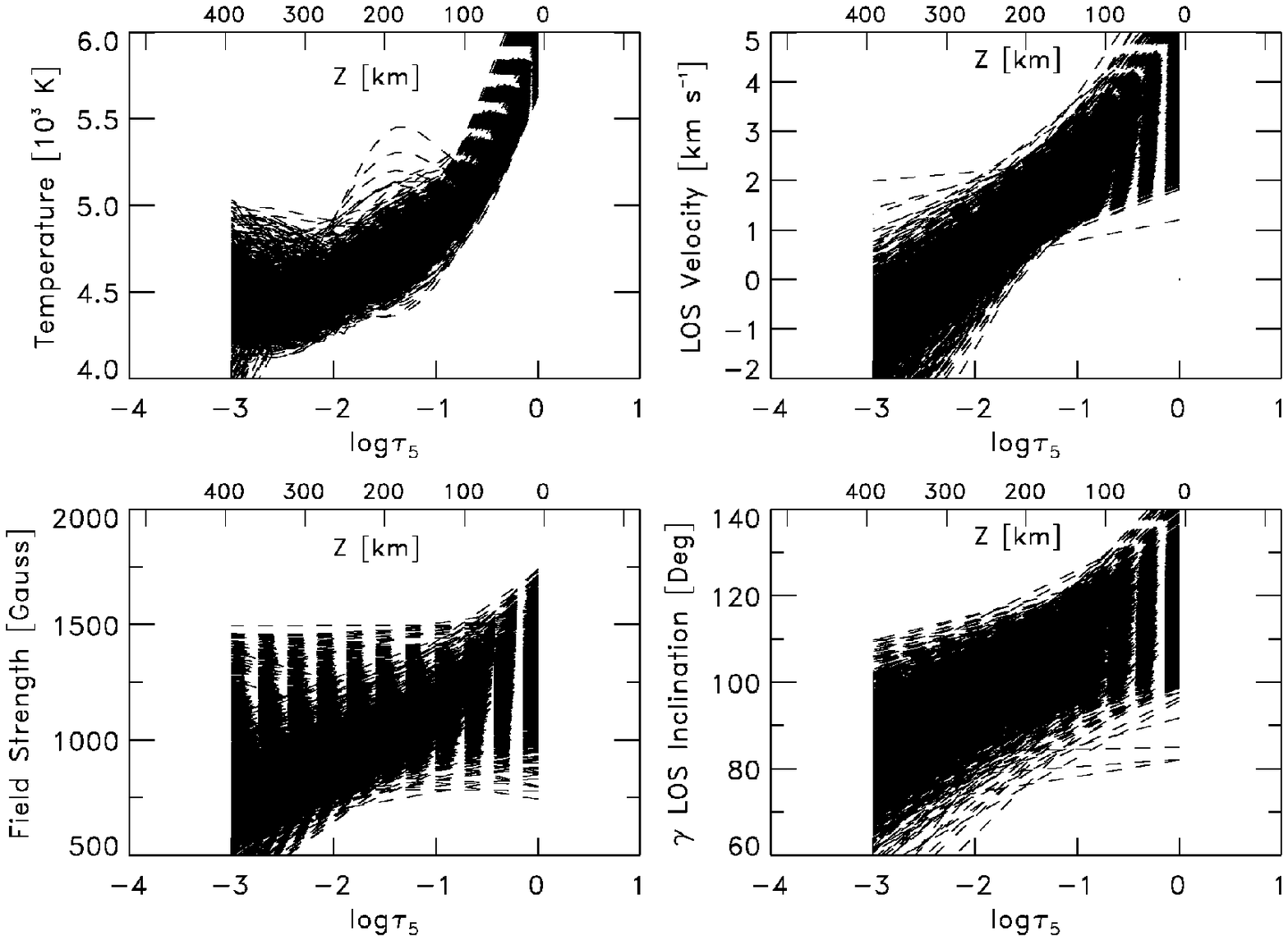}
\figcaption{Same as Fig.~3 but using a {\it local} scattered light profile. About 90 \% of the inverted pixels
show $dB/d\tau >0$.}
\end{center}
\end{figure*}

\begin{figure*}
\begin{center}
\includegraphics[width=14cm]{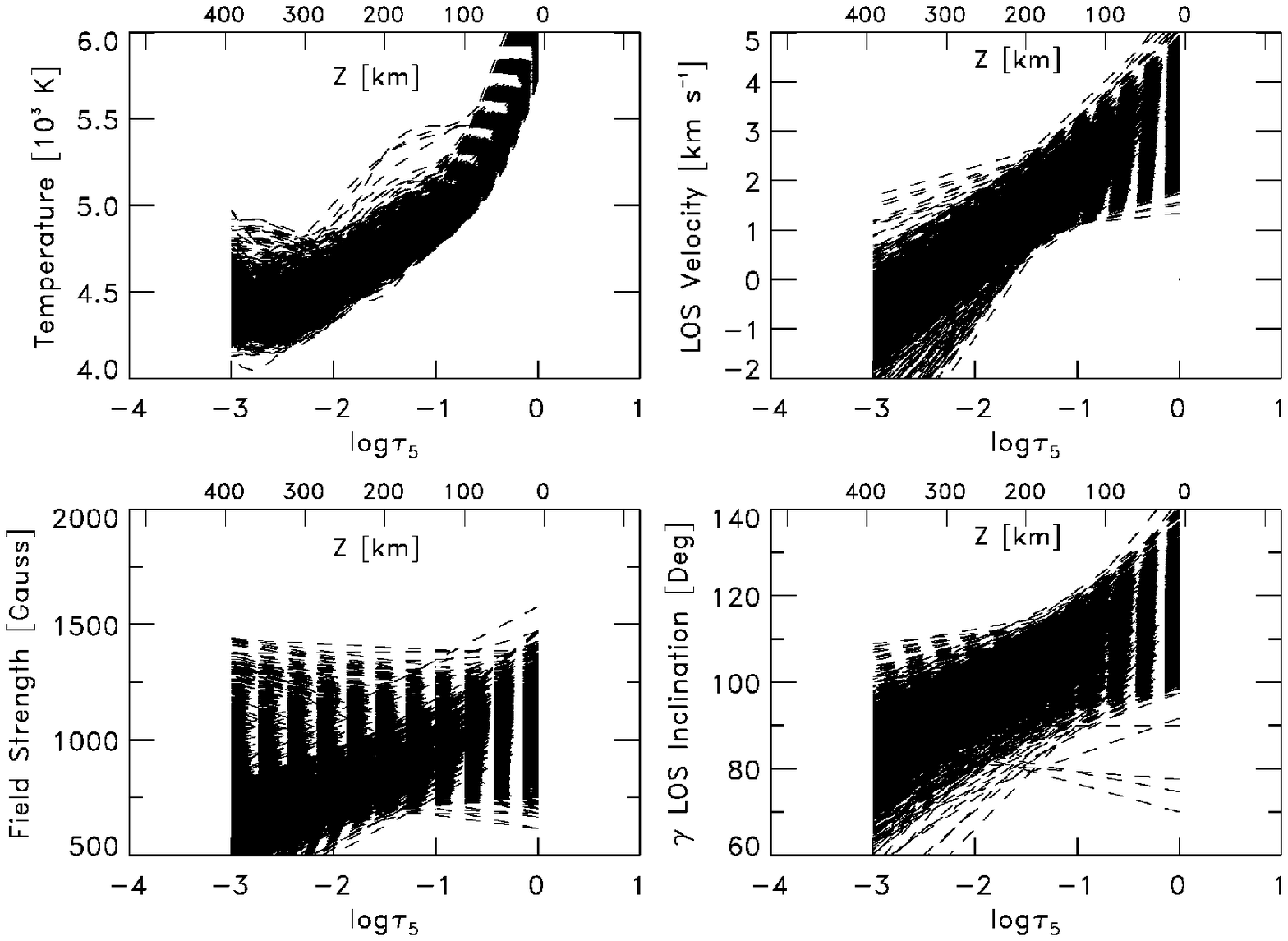}
\figcaption{Same as Fig.~3 but {\it without} the scattered light. About 68 \% of the inverted pixels
show $dB/d\tau >0$.}
\end{center}
\end{figure*}

As a final test, we have repeated our 2-node inversion but neglecting any scattered light: $\alpha_{\rm qs}=0$.
This test is very appropriate because, according to Spruit \& Scharmer (2006), inversions of spectropolarimetric data
fail to detect field-free regions in the penumbra as a consequence of these being already included in the scattered light profile.
If their hypothesis is correct, not accounting for the scattered light contribution should uncover these regions
with $B \simeq 0$ near $\tau_5=1$. Results ignoring the effects of the scattered light (Figure~11)
are essentially unchanged if compared to those where we used a global (Fig.~3) or local (Fig.~10) scattered light profile,
with 68 \% of the pixels showing $dB/d\tau >0$. The only difference is the increased temperatures obtained when we impose  $\alpha_{\rm qs}=0$.
In particular we do not see any pixel where the magnetic field reaches very small values in the deepest photospheric layers.

Taking into account all tests carried out in this section, it seems unlikely that the scattered light can significantly bias our
magnetic field stratifications, consequently making highly unlikely that we are missing the detection of field-free regions near $\tau_5=1$.

\section{Conclusions}%

The {\it uncombed} model postulates that penumbral intraspines are characterized by the presence of horizontal flux tubes embedded 
in a surrounding atmosphere that possesses an inclined magnetic field. According to this model, looking along these regions
should reveal a magnetic field that smoothly decreases at first, but once the flux tube contribution starts, the field strength 
could either increase or decrease. Alternatively, the {\it gappy} penumbral model postulates that instraspines correspond 
instead to regions where convective field-free gaps penetrate the penumbral field. In this case the magnetic field strength should 
also decrease with optical depth at first, but suffer a much larger drop once the line-of-sight 
crosses the field-free gap.

In order to differentiate between these two models, we have used polarimetric data at very high 
spatial resolution, recorded with the spectropolarimeter on-board of the Japanese spacecraft Hinode, to investigate the 
depth variation of the magnetic field strength in the penumbra. We have selected a large number ($\sim$7500) of pixels 
that are representative of weak and horizontal magnetic fields (i.e.: penumbral intraspines) carrying strong
Evershed flows. From the inversion of the Stokes profiles at these locations we find that the magnetic field strength can 
either increase or decrease with optical depth. A more detailed inversion 
of the average Stokes vector over the selected pixels, shows that the magnetic field initially decreases,
between $\log\tau_5 \in [-3,-0.7]$, but increases thereafter until $\log\tau_5 = 0$.

The {\it gappy} penumbral model can explain a smoothly decreasing magnetic field strength
only if the $\tau_5=1$ level is formed above the field free gap, otherwise a much more sudden decrease would 
be observed as the line-of-sight penetrates the field-free plasma. A partial solution to this problem can be found 
if we assume that the gap is not fully evacuated of magnetic field. However, it offers no explanation for about 
66 \% of the selected pixels, where an increasing magnetic field strength with optical depth is observed. The
absence of field-free gaps, as indicated by the inversion, does not in itself imply that there is
no form of convection present in the penumbra, but rather suggests that the convective energy transport takes places
in the presence of a magnetic field (see Zakharov et al. 2008; Rempel \& Sch\"ussler 2008). An example is the roll convection 
proposed by Danielson (1961).

All inferred stratifications are compatible with the scenario proposed by the 
{\it uncombed} model. A magnetic field that decreases smoothly towards the interior
of the Photosphere can be explained by either a flux tube (of any field strength) 
whose upper boundary layer lies below $\tau_5=1$ or, if the upper boundary is above $\tau_5=1$,
with a magnetic field inside the flux tube that is weaker than the magnetic field a
few hundred kilometers above. In addition, a magnetic field that increases towards
the interior of the Photosphere is compatible with a flux tube with an upper boundary layer
above $\tau_5=1$, but with a stronger magnetic field than the one above. This very same configuration can explain a magnetic field that first
decreases and then increases with optical depth, as inferred from the inversion of averaged intraspinal profiles.

We have also studied the effects of the scattered light in our inversions. We have seen that any inaccuracies in its treatment are unlikely
to be a source of error in the stratification of the magnetic field strength. It would be very desirable to make a robust confirmation
of our findings, namely $dB/d\tau > 0$ in the outer penumbra, for a larger number of sunspots at different heliocentric angles
and including also the disk-ward side of the penumbra. A natural extension of this work would be to use the Fe I lines
at 1.56 $\mu$m (which are formed deeper in the Photosphere) to confirm the absence of field-free regions
around $\tau_5=1$. Unfortunately, no such observations exist at the spatial resolution needed to resolve penumbral intraspines ($\simeq$ 0.4"). 
Indeed, some studies at slightly lower resolution (0.6-0.7") have been presented by Cabrera Solana et al. (2008), who used simultaneous observations
of Fe I 630 nm and 1.56 $\mu$m recorded with the TIP (Mart{\'\i}nez Pillet et al. 1999) and POLIS (Schmidt et al. 2003) instruments. They found that
in the outer penumbra, the horizontal magnetic field component (carrying the Evershed flow) was no longer weaker than the more vertical one. 
This can be used as an independent confirmation of our work, where we routinely find $dB/d\tau > 0$ at
large radial distances from the center of the sunspot. In addition, flux tubes with stronger magnetic field than the one of the environment
in which they are embedded are also necessary to explain certain aspects of the net circular polarization observed in the outer penumbra of 
sunspots (Tritschler et al. 2007; Ichimoto et al. 2008).

\acknowledgements{}


\begin{thebibliography}{}
\bibitem[2005]{naza1}
  Bello Gonz\'alez, N., Okunev, O.V., Dom{\'\i}guez Cerde\~na, I., Kneer, F., Puschmann, K.G. 2005,
  A\&A, 434, 317
\bibitem[2005]{luis1}
  Bellot Rubio, L.R., Baltasar, H. \& Collados, M. 2004,
  A\&A, 427, 319
\bibitem[2003]{luis2}
  Bellot Rubio, L.R. 2003, in proceedings of the Solar Polarization conference. Eds: Javier Trujillo and
  Jorge S\'anchez Almeida. ASP Conf. Series, vol. 307, p 301.
\bibitem[2004]{borrero2}
  Borrero, J.M., Solanki, S.K., Bellot Rubio, L.R., Lagg, A. \& Mathew, S.K. 2004,
  A\&A, 422, 1093
\bibitem[2005]{borrero3}
  Borrero, J.M., Lagg, A., Solanki, S.K. \& Collados, M. 2005,
  A\&A, 436, 333
\bibitem[2006]{borrero6}
  Borrero, J.M., Solanki, S.K., Lagg, A., Socas-Navarro, H. \& Lites, B.W. 2006,
  A\&A, 450, 383
\bibitem[2007]{borrero1}
  Borrero, J.M. 2007, A\&A, 471, 967
\bibitem[2008]{borrero5}
  Borrero, J.M., Lites, B.W. \& Solanki, S.K. 2008,
  A\&A, 481, L13
\bibitem[2008]{dani1}
  Cabrera Solana, D., Bellot Rubio, L.R., Borrero, J.M. \& Del Toro Iniesta, J.C. 2008,
  A\&A, 477, 273
\bibitem[1961]{danielson}
  Danielson, R.E. 1961
  ApJ, 134, 289
\bibitem[2003]{josecarlos1}
  Del Toro Iniesta, J.C. 2003.
  Introduction to spectropolarimetry. Cambridge University Press (Cambridge, UK)
\bibitem[2008]{kiyoshi1}
  Ichimoto, K., Tsuneta, S., Suematsu, Y., Katsukawa, Y., Shimizu, T., Lites, B.W., Kubo, M.,
  Tarbell, T.D., Shine, R.A., Title, A.M., Nagata, S., A\&A, 481, L9
\bibitem[2007]{kosugi1}
  Kosugi, T., Matsuzaki, K., Sakao, T. et al. 2007,
  \solphys, 243, 3
\bibitem[2005]{kai1}
  Langhans, K., Scharmer, G., Kiselman, D., L\"ofdahl, M.G. \& Berger, T.E. 2005,
  A\&A, 436, 1087
\bibitem[1993]{bruce1}
  Lites, B.W., Elmore, D.F., Seagraves, P. \& Skumanich, A.P 1993,
  ApJ, 418, 928
\bibitem[2001]{bruce2}
  Lites, Q.W., Elmore, D.F. \& Streander, K.V. 2001, in ASP Conf. Ser. 236,
  Advanced Solar Polarimetry, ed. M. Sigwarth (San Francisco: ASP), 33
\bibitem[1999]{valentin1}
  Mart{\'\i}nez Pillet et al. 1999, in ASP Conf. Ser. 183, High Resolution Solar Physics: Theory, Observations
  and Techniques, ed. T.R. Rimmele, K.S. Balasubramanian \& R.R. Radick (San Francisco: ASP), 264
\bibitem[2000]{valentin2}
  Mart{\'\i}nez Pillet, V. 2000, A\&A, 361, 734
\bibitem[2003]{shibu}
  Mathew, S.K., Lagg, A., Solanki, S.K. et al. 2003,
  A\&A, 410, 695
\bibitem[2007]{david1}
  Orozco Su\'arez, D., Bellot Rubio, L.R. \& Del Toro Iniesta, J.C. 2007a,
  ApJ, 662, L31
\bibitem[2007]{david2}
  Orozco Su\'arez, D., Bellot Rubio, L.R., Del Toro Iniesta, J.C. et al. 2007b,
  ApJ, 670, L61
\bibitem[2008]{matthias}
  Rempel, M. \& Sch\"ussler, M. 2008, {\it in preparation}
\bibitem[1998]{ruedi}
  R\"uedi, I., Solanki, S.K., Keller, C.U. \& Frutiger, C. 1998,
  A\&A, 338, 1089
\bibitem[1992]{basi1}
  Ruiz Cobo, B. \& del Toro Iniesta 1992,
  ApJ, 398, 375
\bibitem[2002]{goran}
  Scharmer, G.B., Gudiksen, B.V., Kiselman, D., Löfdahl, M. G.; Rouppe van der Voort, L.H. M.
  2002, Nature, 420, 151
\bibitem[2006]{spruit2}
  Scharmer, G. \& Spruit, H.C. 2006,
  A\&A, 460, 605
\bibitem[1998]{rolf2}
  Schlichenmaier, R., Jahn, K. \& Schmidt, H.U. 1998,
  A\&A, 337, 897
\bibitem[2003]{wolfgang1}
  Schmidt, W., Beck, C., Kentischer, T., Elmore, D. \& Lites, B.W. 2003,
  Astron. Nach., 324, 300
\bibitem[2007]{shimizu1}
  Shimizu, T., Nagata, S., Tsuneta, S. et al. 2007,
  \solphys, {\it in press}
\bibitem[1993]{sami3}
  Solanki, S.K., \& Montavon, C.A.P. 1993,
  A\&A, 275, 283
\bibitem[2003]{sami4}
  Solanki, S.K. 2003,
  A\&ARv, 11, 153
\bibitem[2006]{spruit1}
  Spruit, H.C. \& Scharmer, G.B. 2006,
  A\&A, 447, 343
\bibitem[1997]{stanchfield}
  Stanchfield, D.C.H., Thomas, J.H. \& Lites, B.W. 1997,
  ApJ, 447, 485
\bibitem[2007]{ali1}
  Tritschler, A., M\"uller, D.A.N, Schlichenmaier, R. \& Hagenaar, H.J. 2007,
  ApJ, 671, L85
\bibitem[1998]{carlosw1}
  Westendorp Plaza, C., Del Toro Iniesta, J.C., Ruiz Cobo, B., Mart{\'\i}nez Pillet, V., Lites, B.W \& Skumanich, A. 1998,
  ApJ, 494, 453
\bibitem[2001]{carlosw2}
  Westendorp Plaza, C., Del Toro Iniesta, J.C., Ruiz Cobo, B., Mart{\'\i}nez Pillet, V., Lites, B.W \& Skumanich, A. 2001,
  ApJ, 547, 1130
\bibitem[2008]{vasily}
  Zakharov, V., Hirzberger, J., Riethmueller, T., Solanki, S.K. \& Kobel, P. 2008, A\&A, {\it submitted}
\end{thebibliography}
\end{document}